\documentstyle[aps,prl,multicol,epsf]{revtex}

\def\bra#1{\langle #1 |}
\def\ket#1{| #1\rangle}

\def\S{{\cal{S}}}
\def\half{{1\over 2}}

\def\R{\hbox{\rm I \kern-5pt R}}

\title{Causality in Time-Neutral Cosmologies}
\author{Adrian Kent} 

\address{Department of Applied Mathematics and Theoretical Physics,
University of Cambridge,\\ 
Silver Street, Cambridge CB3 9EW, U.K.}

\date{18 March 1997, revised 13 August 1998} 
\bigskip

\begin{document}
\maketitle
\begin{abstract}
Gell-Mann and Hartle (GMH) have recently considered  
time-neutral cosmological models in which the initial and 
final conditions are independently specified, and several authors
have investigated experimental tests of such models.  

We point out here that GMH time-neutral models
can allow superluminal signalling, in the sense that it can be 
possible for observers in those cosmologies, by detecting and
exploiting regularities in the final state, to construct devices 
which send and receive signals between 
space-like separated points.   In suitable cosmologies, any
single superluminal message can be transmitted with 
probability arbitrarily close to one by the use of redundant signals.
However, the outcome probabilities of quantum measurements generally
depend on precisely which past {\it and future} measurements take 
place.  As the transmission of any signal relies on quantum 
measurements, its transmission probability is similarly 
context-dependent.  As a result, the standard superluminal
signalling paradoxes do not apply.  Despite their unusual features, 
the models are internally consistent. 

These results illustrate an interesting conceptual point.  
The standard view of Minkowski causality is not an absolutely 
indispensable part of the mathematical formalism of relativistic 
quantum theory.  It is contingent on the empirical observation 
that naturally occurring ensembles can be naturally pre-selected 
but not post-selected.  
\end{abstract}

\begin{multicols}{2}

\newcommand\mathC{\mkern1mu\raise2.2pt\hbox{$\scriptscriptstyle|$}
                {\mkern-7mu\rm C}}                     
\newcommand{\mathR}{{\rm I\! R}} 

\section{Introduction} 

Standard quantum theory is applied in a time asymmetric way.  
The physical state of a quantum system 
is taken to be completely described by the state vector
at any given time, which is taken to be  
derivable from past events. 
The state vector encodes all the available information
about the future behaviour of the system, allowing 
probabilistic predictions about the results of any
sequence of possible future measurements.  

As Aharonov, Bergmann and Lebowitz (ABL) pointed out in their 
classic discussion\cite{abl} of time symmetry and quantum measurement, 
this time asymmetry derives not from the formalism of quantum
theory but from the way we construct the
statistical ensembles to which the theory is applied.  
If ensembles are constructed time symmetrically, using post-selection
as well as pre-selection, then the probability distribution for 
the outcome of any series of experiments at intermediate times is 
time symmetric. 
 
The familiar time asymmetry relies on the 
assumption that ensembles on which measurements have
unambiguous probability distributions can be constructed via 
pre-selection alone.  Every test to date supports
this assumption, but it remains logically
possible that it could be empirically refuted 
without necessarily invalidating the basic formalism of quantum 
theory. 

Recently, Gell-Mann and Hartle (GMH) have 
proposed\cite{gmhthree} a class of time-neutral cosmological models
in which the initial and final quantum states are fixed independently.
These models provide a foil against which to test the standard
time asymmetric understanding of cosmology.  
We describe below a property of GMH 
time-neutral cosmologies which seems to have gone
unnoticed, namely that it can be possible for observers in these
cosmologies to send superluminal signals.  

\section{Time-neutral cosmologies: assumptions and definitions} 

We consider, for simplicity, observers near one end of 
the cosmology, who experience a thermodynamic
arrow of time that is well-defined and unidirectional throughout 
space during their era.  We assume that
the initial conditions are defined by some low entropy state, 
and adopt the observers' time conventions, referring to past and 
future, and to initial and final states, with respect to 
their thermodynamic arrow.  
We suppose that the observers can readily 
use present data to make inferences about past events, and
can readily produce ensembles of quantum systems whose state is 
defined (or at least constrained) by a process of 
pre-selection.  

If the final state is also of low entropy, 
there is no globally defined thermodynamic arrow of time.  
The final state need not be of low entropy, but it must, for the purposes 
of our argument, have some regularities. 
The observers must be able to detect and 
exploit features of the final state which allow them to identify
ensembles of quantum systems that are pre-selected {\it and} 
post-selected in such a way as to alter the standard outcome probabilities 
predicted on the basis of pre-selection alone.   
As we will show, it is easy to construct examples
of time-neutral cosmologies in which this necessary condition holds. 

Any discussion of quantum theory in the context 
of cosmology goes outside the Copenhagen framework,
and so raises difficult interpretational questions.  
GMH interpret their time-neutral cosmological models 
via the consistent histories formulation of quantum 
cosmology.\cite{gmhsantafe}
We focus here on the observational implications for 
standard quasiclassical observers (such as, hypothetically, 
ourselves) in a time-neutral cosmology, which allows us to leave aside most  
interpretational questions.  
Instead of considering all the infinitely many incompatible
consistent sets defined by GMH, we suppose, as part of the 
definition of a time-neutral 
cosmological model, that one particular quasiclassical 
consistent set $\S$ has been fixed.  All statements about
quasiclassical events --- the actions of observers, their manipulations
of their experimental apparatus, and their observations --- are
statements which should in principle be understood in terms of 
quasiclassical variables defined by that fixed 
set.\footnote{We assume here familiarity with the consistent histories
approach to quantum cosmology. 
For reviews, attempts at a definition of quasiclassicality, and discussions
of the problems see, for example, 
Refs. \cite{gmhsantafe,gmhprd,dowkerkentone,dowkerkenttwo}.}

So, the full cosmological theory is defined 
by initial and final states, $\rho_i$ and $\rho_f$, which  
are non-orthogonal positive semi-definite matrices, and 
the specified quasiclassical consistent set of histories
$\S$.  We implicitly require also some theory of the 
hamiltonian and specification of canonical variables which allows us 
to make physical sense of the events abstractly 
specified in $\S$ --- allowing us, for example, to interpret
suitable projections $P$ on Hilbert space in terms of local number 
density operators integrated over small regions, and so to 
interpret the statement that $P$ is realised as the statement 
that the density $d$ of matter in a certain specified volume lies
in the range $d_{\rm min} \leq d \leq d_{\rm max}$.  

The probabilities 
for particular quasiclassical events --- for example, a given
observer pressing the switch of a signalling device --- are now, in
principle, calculable from the boundary conditions by the 
standard decoherence functional probability formula.\cite{gmhsantafe}  

A comment on free will is necessary at this point.
When we press a switch, it generally seems 
to us that we could have chosen not to.\footnote{Whether or not this
impression of free choice is justifiable is, of 
course, beyond the scope of this paper!}
If our pressing a switch is reliably correlated with someone else 
receiving a message --- in such a way that we can give a 
direct physical explanation of the fact
that the message will be received if and only if we 
press the switch --- then we describe ourselves as sending a 
signal by the action of pressing the switch.  
This is the operational definition used in 
this paper when, for example, we speak about 
an observer {\it choosing} to send a signal. 
We will show that time-neutral
cosmological models in which the final state has a detectable 
regularity supply a mechanism by which intelligent observers 
can construct devices which allow them --- in this operational
sense --- to send superluminal signals.   

One can imagine pathological cosmological models in which, via 
some conspiracy of the boundary conditions, it is pre-determined 
that observer A will press a switch a number of times, and that 
observer B will receive a message at correlated times, although 
there is no direct physical link between the switch and the message.  
Such coincidences have no simple physical explanation: they can 
{\it only} be understood from a detailed analysis of the 
boundary conditions.  It cannot then be said that pressing the 
switch causes the message to be received, in the conventional sense.  
As will be seen, pathologies of this sort are not invoked here.  

We suppose that the specific quasiclassical events to be  
discussed occur in a small, flat region of space-time in
roughly our own cosmological era --- near the earth, in the 
near future, for example. 
We suppose too that (as is generally believed) the 
simplest non-relativistic version of 
the consistent histories formalism adequately  
describes localised quasiclassical physics: i.e., we 
take physical events to be 
defined by projections at given times in some fixed frame,

That is, we assume that some partial and coarse-grained
quasiclassical history $H$, built from basic events in $\S$, and
sufficient to describe the evolution of the
classical structure of the universe up to the present,
has already been realised. 
All subsequent probabilities should now, in principle, be
calculated by conditioning on the realisation of $H$.   
To avoid the need for this, we consider events defined by
interactions of pre- and post-selected quantum subsystems with 
classical measuring devices, and assume that the 
subsystems are uncorrelated with the quasiclassical degrees 
of freedom at all times between the initial condition and their 
first measurement, and again between their final measurement 
and the final condition.  

\section{Superluminal signalling} 

A single example suffices to show that 
time-neutral cosmologies are not immune from superluminal
signalling.  It is hard to make the example very realistic, 
for several reasons.  First, neither quantum cosmology nor 
quantum gravity are well understood. 
Second, there is no precise mathematical definition of 
quasiclassicality.  
Third, no detailed proposals for plausible time-neutral cosmologies 
have been made.  Indeed, while 
time-neutral quantum theory itself is certainly consistent,
it is less clear that time-neutral cosmological models with
the specific properties that GMH consider\cite{gmhthree} are 
consistent with present theories and 
observations.\footnote{A thoughtful discussion of some of the 
problems that arise in defining GMH time-neutral 
cosmologies can be found in Ref. \cite{craig}.} 
We can, however, produce an example of superluminal signalling
under assumptions which, while contrived, are not excluded by any 
theoretical principle.  

Suppose that the cosmology can be divided into two
subsystems, which interact only briefly.  
The first subsystem includes an entire quasiclassical
realm, containing observers, their measuring devices, and all the
large-scale cosmological structure.  It is to be described by
some consistent set $\S$, which is sufficiently detailed to 
describe the actions of the observers and the experiments they
carry out.  We can assume that this realm, the observers and their 
experimental devices are very like our own, us and ours.  

The second subsystem is a large supply of pairs of initially entangled
particles initially 
in a singlet spin state 
$$ \ket{ B} =
1/\sqrt{2}( \ket{\uparrow}_L \ket{\downarrow}_R - \ket{\downarrow}_L
\ket{\uparrow}_R ) \, $$
where the states $\ket{\uparrow}$ and
$\ket{\downarrow}$ are the $\pm \half$ eigenstates of $\sigma_z$, for  
a choice of the $z$ axis which is unambiguously defined by parallel
transport throughout their propagation.  
These pairs are otherwise initially 
uncorrelated with each other and with the rest of the quasiclassical
realm. 

The pairs of particles propagate freely
from the beginning of the cosmology, undergoing no significant interactions, 
with the L and R particles propagating in distinct directions, 
until they arrive at two regions populated by experimentalists. 
The experimentalists have set up measuring 
devices, which register the arrival of each particle and 
measure its spin, about some axis or other --- the axes are 
initially altered quite often by the experimentalists, as they 
attempt to investigate the ensemble statistics for various
choices of axes.  Their measurements are idealised 
von Neumann measurements, so that 
the standard ABL time-neutral probability formula applies.

The particles arrive sequentially pair by pair: i.e  
the first particles to arrive at each apparatus are 
entangled, as are the second and every successive pair. 
These arrival events have a large spacelike separation. 
After a single measurement, they propagate freely off into
empty space, never again undergoing any significant interaction 
until the end of the cosmology.   
At this point, they satisfy the final condition that the R 
particles are found, with certainty, in the state $\ket{\uparrow}_R$;
the final state of the L particles is not constrained.

The full initial state for the cosmology is thus 
$\rho_i \otimes \rho_B \otimes \cdots \otimes \rho_B$, where 
$\rho_B = \ket{B} \bra{B}$ describes an initially Bell-entangled
pair.
The density matrix $\rho_i$ describes the initial conditions 
from which the rest of the quasiclassical realm emerges.
The full final state, similarly, is 
$\rho_f \otimes \rho' \otimes \cdots \otimes \rho'$, 
where $\rho' = I_L \otimes \ket{\uparrow}_R \bra{\uparrow}_R$, 
and $I_L$ is the identity operator on the left hand particle spin. 
In both cases, the other degrees of freedom describing the initial
states have been omitted.
The precise choice of the density matrix $\rho_f$, which
describes the final state of the rest of the quasiclassical realm,
is unimportant.  

The experimentalists exchange all their data via
classical signals, and search for correlations in those data.
Familiar with ABL quantum theory, they eventually correctly conclude 
that the pairs they observe are pre- and post-selected to arrive in  
the state $\rho_B$ and to reach the final state $\rho'$.  

As they realise, these conditions mean that the 
outcome probabilities for the experimenters on the left
depend on the axes chosen by the experimenters on the right. 
For instance, if the experimenters on the right choose 
the $z$ axis, the post-selection implies that they will,
with probability one, obtain the eigenvalue $+\half$.  
If the experimenters on the left also choose
the $z$ axis for the corresponding particle, they will also,
with probability one, obtain the eigenvalue $-\half$.  
On the other hand, if the experimenters on the right choose
the $x$ axis, they will have probability $\half$ of finding their
particle in either of the possible spin eigenstates. 
The left hand experimenters, if they still choose the $z$ axis, 
will then also have probability $\half$ of finding the corresponding
particle in either eigenstate. 

The experimenters now 
use classical signals to agree a superluminal signalling protocol
for future use.  For the right hand experimenters to send 
a $0$, they choose the $z$ 
axis for their observations on an agreed sequence of $m$ particles.  
To send a $1$, they choose the $x$ axis for 
the same sequence of particles.
The left hand experimenters always choose 
the $z$ axis for their observations.  If they obtain the 
eigenvalue $-\half$ for each of the $m$ particles, they
record a $0$; if not, they record a $1$. 

This is clearly a probabilistic (and not optimally efficient)
protocol: the probability, $2^{-m}$, of misreading a $1$ as a $0$
is small for large $m$, but cannot be made to vanish.
Nonetheless, if the spacelike separation is large compared with 
the interval between the arrival of successive particles, the
protocol allows long messages to be sent superluminally 
and very reliably.  
In information theoretic terms,\cite{shannonweaver}
the information transmitted superluminally
is non-zero and can be made arbitrarily close to one bit by taking
$m$ large.  Standard causality is thus violated. 

This signalling protocol 
will work only under the specified conditions.
Actions of the experimenters or later observers could frustrate it.  
If the particles re-interact with the 
quasiclassical world after the first measurement, the outcome 
probabilities will be altered, and the signalling will not
necessarily work.   

There is no contradiction here, however.  The assumptions made 
in the example correspond to a possible cosmology.  If, according 
to the initial and final conditions of that cosmology, the 
experimenters are bound --- in all histories of the set $\S$ --- to carry out
precisely one observation on each particle in each pair, and 
to allow the particles to propagate unhindered, then the 
protocol will work.  If these assumptions hold true in some histories
of the set, the protocol will work if the realised history is one
of those, and will not necessarily work otherwise.  
There is no reason why it should be true, in a general
cosmology, that the probability of the assumptions being satisfied
is large, or even non-zero. 
However, we may (and do) consider some realised history in
some particular cosmology in which they do, in fact, hold. 

Whatever the details 
of the cosmology, any time-neutral theory of GMH
type, together with the specification of some consistent 
quasiclassical set, will unambiguously assign probabilities 
to possible histories in a logically consistent way.
Thus, although the theories allow superluminal signalling, 
there is no possibility of setting up a paradoxical causal loop in a way
which leads to an internal contradiction: some internally consistent
outcome will be realised.  
In standard superluminal signalling paradoxes,  
a signaller can send signals into his own past light cone, 
and one can imagine things being arranged so that the subsequent 
transmission of the signals then becomes impossible --- by, for
example, killing the signallers' grandfather as a 
child.\cite{rindler} 
Such paradoxes break down here 
because the probabilistic predictions in a time-neutral
model are context-dependent.  
Superluminal signalling devices can have efficiency arbitrarily 
close to one in isolation, but if a paradoxical causal loop 
is set up involving such devices, at least one device
will fail to function --- not through any mysterious intervention 
or because of any ad hoc postulate, but as a result of 
the standard probability rules of the consistent histories
formalism.  

Nor is there any violation of Lorentz invariance in any standard 
sense: the theory itself picks out no preferred frame, and its
predictions can equally well be calculated in any frame.  

\section{Conclusions} 

We have seen that GMH time-neutral cosmologies can 
allow superluminal signalling, given knowledge of a detectable
regularity in the final state --- knowledge which can be 
empirically obtained by observers in those cosmologies. 

The example given is admittedly contrived. 
However, the possibility of superluminal signalling in these 
cosmologies is more general.  
Any final state with
interesting regularities is liable to allow superluminal signalling.
For the above argument to go through, all that is required 
is that, at some point in a realised history, ensembles of 
entangled pairs of quantum subsystems can be produced in which at
least one member of the pair, if left to propagate freely, would,
as a result of post-selection, have a propensity to end up in some 
particular one of the entangled eigenstates.  

For example, the time-neutral cosmologies GMH discuss
involve the formation of large-scale structure at
both ends of the universe.   It is, admittedly, hard to prove 
any rigorous result about the properties of such cosmologies.
However, if they can actually be realised, it seems clear 
that, from the perspective of observers nearer (what they regard as) 
the initial state, there should be a propensity for matter 
eventually to propagate into the stars and galaxies formed in
the time-reversed sense at the other end of the cosmology, in a 
way which eventually must result in a detectable
deviation from the predictions of the time asymmetric quantum theory
defined by the initial state alone.  
It should then be possible to use this propensity to construct
superluminal signalling devices of the type described above, 
by using pairs of particles entangled in position space. 

Paradox is avoided as the probability
that any signal is successfully transmitted is context-dependent. 
Signalling devices, however reliable when operated in isolation,
will generally become unreliable if any paradoxical loop is attempted. 
Though superluminal signalling is usually seen as a fatal flaw, 
it causes no inconsistency in time-neutral cosmologies.

Of course, there are independent  
reasons to approach GMH time-neutral cosmologies with scepticism. 
The hypothesis that initial causes suffice 
seems very well established.  
Also, the evidence appears to lean against 
a closed universe, and a closed universe is the natural cosmological 
setting for time-neutral models, 

Still, it seems clear that the reason why standard
quantum theory respects Minkowski space causality --- in the
sense that it does not allow superluminal signalling ---
is not quite as usually suggested.  
It is not absolutely necessary to forbid superluminal signalling
in order to maintain the internal consistency of a relativistic
quantum theory.  
The impossibility of superluminal signalling is, rather, 
ultimately a hypothesis, contingent on the fact that naturally
occurring ensembles can be identified on the basis of pre-selection
alone.  

Price has recently suggested\cite{price} that backward and 
forward causation could in principle simultaneously be present 
in a well-defined physical theory.  It follows from
the discussion above that 
this is indeed a property 
of suitable time-neutral cosmological theories. 
To explain the sense in which this is so, we need first
to say something about the language of causality in
standard time asymmetric theories.\footnote{For a detailed 
discussion of causality the reader is enthusiastically referred to 
Ref. \cite{price}.  The view sketched in the following brief 
comments is indebted to, but not necessarily endorsed by, 
and certainly not an adequate summary of, Price's discussion.} 

As mentioned above, according to essentially all 
current physical theories, our actions are in principle 
probabilistically predictable from the boundary conditions.  
Nonetheless, it is generally regarded as useful to employ
the language of causality in describing of our actions --- to say,
for example, that we are able to causally influence the future,
but not the past.  
Various justifications are given.
It is sometimes intended that causes and effects
are to be distinguished by the asymmetry of their correlations. 
Many independent future effects are generally correlated 
with, and can be traced to, a single cause.  If, for
example, we send an ordinary radio signal, it may be received
and acted upon by many people in the future.  
Causal statements about the consequences of our own actions can
also be meant simply as statements about our own
perspective on the world.  It appears to us that we have a free
choice to send the signal or not, and that 
the events which follow from our sending the signal
are contingent on and caused by our exercising the choice to send it.  
 
In either of these conventional senses, when time-neutral 
cosmological theories allow superluminal signalling, they generally  
allow bidirectional causation.\footnote{More generally, the claim is 
that there is no sensible way of maintaining the language of
causation, with respect to the actions of agents like ourselves,
in GMH time-neutral theories, without admitting 
the possibility of bidirectional causation.}   
Agents in those theories who, 
from their perspective, have evolved to causally influence the 
future, can also causally influence the past, for example by setting up
two or more superluminal signalling devices connected in such a way
as to send signals into their past light cone --- when, that is,
the boundary conditions allow such constructions, and 
some boundary conditions certainly do. 
Again, though, the past can never be influenced in a way which will lead
to contradictions --- it is always impossible, for example, for
agents to undo their present observations by acting on the past.  

Finally, the fact that time-neutral theories allow superluminal 
signalling may have interesting implications for discussions of 
the properties of other non-standard variants of quantum theory.
It is often suggested that non-relativistic theories which allow superluminal 
signalling --- for example, non-linear generalisations 
of quantum mechanics --- cannot possibly 
have a consistent relativistic extension.  
While this may indeed be true in specific cases of interest,  
some further argument is clearly needed: the demonstration of 
superluminal signalling is not sufficient {\it per se}.   

Of course, one can always argue that theories which allow 
superluminal signalling should be rejected on grounds of 
aesthetics or by Ockham's razor.  The discussion above, which is 
solely concerned with the logical relations among physical
hypotheses, provides no counter-argument here. 
\vskip15pt
\noindent{\bf Acknowledgements}  
I am grateful to colleagues at York University, Cambridge
University, King's College, London, and Oxford University 
for helpful comments in recent seminars on this topic, and to 
the Royal Society for financial support. 


\end{multicols}

\end{document}